\documentclass[aps,pra,amsmath,onecolumn,amssymb,footinbib,superscriptaddress,showpacs,notitlepage]{revtex4-1}

\usepackage[T1]{fontenc}
\usepackage[latin9]{inputenc}
\usepackage{color}
\usepackage{graphicx,epstopdf}
\usepackage{url}
\usepackage[normalem]{ulem}
\usepackage{amsthm}
\usepackage{amsmath}
\usepackage{hyperref}

\def\be{\begin{equation}}
\def\ee{\end{equation}}
\def\bea{\begin{eqnarray}}
\def\eea{\end{eqnarray}}
\def\bi{\begin{itemize}}
\def\ei{\end{itemize}}
\def\bin{\begin{enumerate}}
\def\ein{\end{enumerate}}
\def\la{\langle}
\def\ra{\rangle}

\newcommand{\vect}[1]{\mathbf{#1}}

\begin{document}

\title{Efficient algorithm to compute the second Chern number in four dimensional systems}

%%%%%%%%%%%%%%%%%%%%%%%%%%%%%%%%%%%%%%%%%%%%%%%%%%%%%%%%%%%%%%%%%%%%%%%%%%%%%%%
\author{M. Mochol-Grzelak}
\affiliation{
Instytut Fizyki imienia Mariana Smoluchowskiego and
Mark Kac Complex Systems Research Center, 
Uniwersytet Jagiello\'nski, ulica prof. Stanis\l{}awa \L{}ojasiewicza 11, PL-30-348 Krak\'ow, Poland}
\affiliation{ICFO-Institut de Ciencies Fotoniques, The Barcelona Institute of Science and Technology, 08860 Castelldefels (Barcelona), Spain}
\author{A. Dauphin}
\email{alexandre.dauphin@icfo.eu}
\affiliation{ICFO-Institut de Ciencies Fotoniques, The Barcelona Institute of Science and Technology, 08860 Castelldefels (Barcelona), Spain}
\author{A. Celi}
\affiliation{Center for Quantum Physics, University of Innsbruck, and Institute for Quantum Optics and Quantum Information, Austrian Academy of Sciences, Innsbruck, Austria} 
\affiliation{ICFO-Institut de Ciencies Fotoniques, The Barcelona Institute of Science and Technology, 08860 Castelldefels (Barcelona), Spain}
\author{M. Lewenstein}
\affiliation{ICFO-Institut de Ciencies Fotoniques, The Barcelona Institute of Science and Technology, 08860 Castelldefels (Barcelona), Spain}
\affiliation{ICREA-Institucio Catalana de Recerca i Estudi Avan\c cats, E-08010 Barcelona, Spain}

\date{\today}

\begin{abstract}
Topological insulators are exotic material that possess conducting surface states protected by the topology of the system. They can be classified in terms of their properties under discrete symmetries and are characterized by topological invariants. The latter has been measured experimentally for several models in one, two and three dimensions in both condensed matter and quantum simulation platforms. The recent progress in quantum simulation opens the road to the simulation of higher dimensional Hamiltonians and in particular of the 4D quantum Hall effect. These systems are characterized by the second Chern number, a topological invariant that appears in the quantization of the transverse conductivity for the non-linear response to both external magnetic and electric fields. This quantity cannot always be computed analytically and there is therefore a need of an algorithm to compute it numerically. In this work, we propose an efficient algorithm to compute the second Chern number in 4D systems. We construct the algorithm with the help of lattice gauge theory and discuss the convergence to the continuous gauge theory. We benchmark the algorithm on several relevant models, including the 4D Dirac Hamiltonian and the 4D quantum Hall effect and verify numerically its rapid convergence.
\end{abstract}
\maketitle

%%%%%%%%%%%%%%%%%%%%%%%%%%%%%%%%%%%%
\section{Introduction}
%%%%%%%%%%%%%%%%%%%%%%%%%%%%%%%%%%%%

Topological insulators and topological states of matter have attracted much interest in the last decades~\cite{Qi_2011,Hasan_2010}. These exotic phases possess physical properties, like the electric conductivity, that rely only on a global feature of the system and are therefore robust with respect to local perturbations. This robustness arises from the topological protection of such systems. The latter can be symmetry protected as in topological insulators, or originates from a global topological order, a new phase of matter that cannot be identified through local order parameters, with a robust degeneracy of the groundstate and global excitations of non-trivial statistics, that can be exploited for realizing quantum memories and for quantum computation~\cite{pachos_2012}.
Specifically, topological insulators are remarkable materials in which currents are carried by surface states protected by the topology of the system~\cite{Qi_2011,Hasan_2010}. A classification of the topological insulators without interactions has been realized in terms of their properties under discrete symmetries in the celebrated \emph{``periodic table''} of topological insulators~\cite{Ching-Kai2016} and a great effort has been done recently to study systems that go beyond the periodic table such as for example topological Mott insulators~\cite{Raghu_2008}, Floquet topological insulators~\cite{Kitagawa_2010}, time crystals~\cite{Sacha2018}, crystalline topological insulators~\cite{Fu_2011}, Weyl semimetals~\cite{Wan_2011}, non-Hermitian systems~\cite{gong2018topological} and the topology in quasicrystals~\cite{Kraus:2016aa,Tran2015,Dareau2017,Fuchs2016}. 

Topological insulators have been observed in condensed matter experiments in two dimensions such as the quantum Hall effect~\cite{Klitzing:1980p1353} and the quantum spin Hall effect~\cite{K_nig_2008,Konig_2007} as well as in three dimensions~\cite{Hsieh_2008}. Complementary to these experimental realizations, quantum simulation of topological phases has been performed in cold atoms~\cite{Aidelsburger_2013,Miyake_2013,Aidelsburger_2014,Jotzu_2014}, photonic crystals~\cite{Rechtsman_2013,Mittal_2016}, photonic quantum walks~\cite{Kitagawa_2012,Cardano_2016}, and even classical systems~\cite{Huber2016}. In particular, cold atoms allow one to simulate topological phases in a very clean and controllable environment~\cite{lewenstein2017,goldman2014,aidelsburger2017artificial,cooper2018topological}. The generation of the artificial gauge field is very challenging and gave rise to a plethora of proposals such as laser assisted tunneling~\cite{Jaksch_2003,Osterloh2005}, shaking~\cite{Eckardt2005,Hauke2012}, synthetic dimensions~\cite{boada_2012,Celi_2014}. Recently, the quantum simulation of several models of topological insulators has been experimentally realized using these or related techniques: e.g. the Hofstadter model with bosons~\cite{Aidelsburger_2013,Miyake_2013,Aidelsburger_2014} and the Haldane model for fermions~\cite{Jotzu_2014}, with a combination of laser assisted tunneling and shaking, and the 1D SSH model~\cite{Meier_2016} the Hofstadter strip for bosons ~\cite{Stuhl_2015} and fermions~\cite{Mancini1510} with the help of synthetic dimensions. Note that these narrow systems in appropriate conditions display both edge states excitations and topological response to charge pump as the large ones~\cite{Mugel2017}. The topological insulators and their associated topological invariants have been measured in in several experiments. In one dimension, the winding number and the Zak phase have been measured in cold atoms in superlattices~\cite{Atala_2013}, in photonic systems~\cite{Zeuner2015,Cardano_2016}, in superconducting circuits~\cite{Flurin2017}  and the robustness with respect to disorder has been recently probed in disordered atomic wires~\cite{Meier2018}. The Chern number has also been measured  through pumping experiments in 1D superlattices~\cite{Lohse_2015,Nakajima_2016} and its corresponding boundary phenomena were optically probed in 1D quasicrystals~\cite{Kraus2012,verbin2015}. In two dimensions, the Chern number has been measured through the dynamics of the center mass of the atomic cloud~\cite{Aidelsburger_2014}, through the measurement of the depletion rate in the presence of an external driving~\cite{Asteria_2018} and through the dynamics of highly excited states of the system~\cite{Tarnowski2017}. Furthermore, the geometrical properties of the bands such the berry curvature has been measured in two recent experiments~\cite{Li2016,Flaschner_2016}.

Current experiments are now going beyond one, two and three dimensions. Topological insulators in higher dimension exist and are especially interesting as they display novel features and responses that are characterized by novel topological invariants. For instance, the generalization of the 2D quantum Hall effect to four dimensions was proposed~\cite{Frohlich_2000,Zhang_2001}. There, the magnetic field is replaced by a SU(2) gauge field. The main new feature of such model is that Hall response becomes quadratic in the applied electric field and proportional to the second Chern number, which is trivially zero in topological models in lower dimensions. Simulating such model amounts to realize a 4D Dirac system that can be achieved for instance with synthetic dimensions or by mapping it to a 1D dynamical system~\cite{Edge2012}. A simpler incarnation of the 4D quantum Hall can be obtained from the product of two separate 2D-quantum Hall systems. That is, the two (Abelian) magnetic fields are applied in two perpendicular but not intersecting planes. This model displays linear and quadratic responses to applied external fields, with the former proportional to the first Chern number as in 2D quantum Hall effect,  and the latter to the second Chern number. The simulation of this second version of the 4D quantum Hall effect was first proposed in a quasi-crystal in two dimensions~\cite{Kraus_2013}.  An alternative way to achieve this model was recently proposed with the help of synthetic dimensions~\cite{Price_2015,Ozawa_2016}.  Very recently, the 4D quantum Hall effect have been observed by using a 2D Thouless pump with ultracold atoms in 2D optical superlattice~\cite{Lohse2018} and its corresponding boundary phenomena were optically probed in a 2D tunable waveguide~\cite{Zilberberg2018}. The second Chern number has also been measured in the quantum simulation of a non-Abelian monopole~\cite{Sugawa2018}.

In the aforementioned models, the second Chern number can always be computed in terms of a product of first Chern numbers but in general this is not case. There, the calculation of the second Chern number can be challenging, especially in the case where the Hamiltonian cannot be diagonalized analytically. In this work, we propose and construct an efficient algorithm to compute the second Chern number. Our algorithm is a generalization of the one proposed by Fukui, Hatsugai and Suzuki~\cite{Fukui_2005} and is based on lattice gauge theory~\cite{Wilson_1974,Phillips_1985}. We illustrate the different steps of the algorithm, and we discuss the equivalence to the continuous gauge theory and its convergence in terms of the grid parameter $\Delta$. We then benchmark the algorithm on several paradigmatic examples. The first one is the lattice Dirac model~\cite{Qi_2008}. This model has the advantage that it can be solved analytically and allows one to compare the results of the algorithm with the exact results. We find numerically the convergence in terms of the grid parameter and show that it corresponds to the theoretically predicted one. The second model is the 4D quantum Hall system obtained from the product of two 2D-quantum Hall ones. This model is experimentally relevant since the computation of the second Chern number is necessary to characterize what could be observed in experiments. Finally, we study a modified version of the 4D quantum Hall effect where the model cannot be factorized anymore in a product two 2D-quantum Hall effect. In this case, there is a need for an efficient algorithm to compute numerically the second Chern number. 

The structure of the paper is presented as follows:

\begin{enumerate}
\item In section II, we review the definitions of the Chern number and the second Chern number in the continuum.
\item In section III, we outline the different steps of the algorithm. We start to review the construction of the lattice gauge theory in two dimensions, introduce the lattice Chern number, and characterize the convergence to the continuous lattice gauge theory. We then generalize the algorithm to four dimensions and discuss its convergence to the continuum.
\item In section IV, we benchmark the algorithm on three paradigmatic examples: the lattice Dirac Hamiltonian, the 4D quantum Hall effect and the generalized 4D quantum Hall effect.
\end{enumerate}

%%%%%%%%%%%%%%%%%%%%%%%%%%%%%%%%%%%%
\section{Topological invariants}
%%%%%%%%%%%%%%%%%%%%%%%%%%%%%%%%%%%%

We here review topological invariants related to the quantum Hall effect. In two dimensions, the relation between the the quantization o the transverse conductivity and the Chern number was shown by Thouless {\it et al.} within the linear response theory~\cite{Thouless_1982}. In four dimensions, the relation between the non-linear response to the current and the second Chern number was shown by Zhang and Hu~\cite{Zhang_2001}.This response has been very recently observed experimentally in Refs.~\cite{Lohse2018} and~\cite{Zilberberg2018} in a 2D Thouless pump experiment.

%%%%%%%%%%%%%%%%
\subsection{First Chern number}
%%%%%%%%%%%%%%%%
In two dimensions, the linear response of the current under an external electric field $\mathbf{E}$ is proportional to the topological invariants of the energy bands. For a Fermi energy $\varepsilon_F$ in an energy gap, the transverse current is quantized and proportional to the total Chern number $C_1(\varepsilon_F)$ of the occupied energy bands~\cite{Thouless_1982}:

\be
\label{eq:current2D}
\begin{split}
j_l&=\frac{e^2}{h} C_1(\varepsilon_F)\epsilon_{lm} E_m,\\
C_1(\varepsilon_F) &= \sum_{\varepsilon_\alpha<\varepsilon_F} \frac{1}{2\pi i}\int_{BZ}F^\alpha_{xy}(\vect{k})d^2k.
\end{split}
\ee

$F^\alpha_{xy}(\vect{k}) = \partial_{k_x}A^\alpha_y(\vect{k})-\partial_{k_y}A^\alpha_x(\vect{k})$  is the Berry curvature and $A^\alpha_{l}(\vect{k}) = \la\psi^\alpha(\vect{k})|\partial_{l}|\psi^\alpha(\vect{k})\ra$ is the Berry connection. Here, $\vert \psi^\alpha(\mathbf{k})\rangle$ denotes the eigenstate of the band $\alpha$ and of quasi-momentum $\mathbf{k}$. The Chern number is an integer and is invariant as long as the energy gap does not close. In the case of degenerate energy bands, the definition of the total Chern number has to be generalized to~\cite{nakahara}:

\be
\label{eq:firstchernnum}
C_1(\varepsilon_F) =  \frac{1}{2\pi i}\int_{{BZ}}d^2k \; \textrm{Tr}[F_{xy}(\mathbf{k})],
\ee

where the non-Abelian Berry curvature $(F_{xy}(\vect{k}))^{\alpha\beta} = \partial_{x}A^{\alpha\beta}_y(\vect{k})-\partial_yA^{\alpha\beta}_x(\vect{k}) + i[A_{x},A_{y}]^{\alpha\beta}$ is written in terms of the Berry connection {of the occupied bands} $(A_\mu(\vect{k}))^{\alpha\beta} = \la \psi^\alpha(\vect{k})|\nabla_\mu|\psi^\beta(\vect{k})\ra$ and the trace is taken over the occupied bands.

%%%%%%%%%%%%%%%%
\subsection{Second Chern number}
%%%%%%%%%%%%%%%%

In four dimensions, the response of the current to an electric field $\mathbf{E}$ and a magnetic field $\mathbf{B}$, which in $D=4$  is defined as the dual of spatial component field strength and is a tensor with $D-2=2$ indices, is related to two topological invariants: the linear response is related to the total Chern number and the non-linear response to the total second Chern number. For a Fermi energy $\varepsilon_F$ lying in an energy gap, the response to external electric and magnetic fields can be written as~\cite{Zhang_2001}:

\begin{align}
&j_l=\frac{e^2}{h}\sum_{\varepsilon_\alpha<\varepsilon_F}\frac{1}{(2\pi)^4}E_m \int_{BZ}F^\alpha_{lm}\,d^4k+\frac{C_2(\varepsilon_F)}{4\pi^2}\epsilon_{lmno}\,E_m\, B_{no}\text{,}\cr
&C_2(\varepsilon_F)=\sum_{\varepsilon_\alpha<\varepsilon_F}  \frac{1}{32\pi^2}\int_{BZ} d^4k\;\epsilon_{lmno}\,F^\alpha_{lm}(\mathbf{k})F^\alpha_{no}(\mathbf{k}). \label{eq:current4D}
\end{align}

The linear response is proportional to the integral on the Brillouin zone of the Berry curvature $F^\alpha_{lm}(\vect{k}) = \partial_{k_l}A^\alpha_m(\vect{k})-\partial_{k_m}A^\alpha_l(\vect{k})$ written in terms of the Berry connection $A^\alpha_{l}(\vect{k}) = \la\psi^\alpha(\vect{k})|\partial_{l}|\psi^\alpha(\vect{k})\ra$. The non-linear response is proportional to the total second Chern number written in terms of the second Chern number of the band $\alpha$.

In the case of degenerate bands, the total second Chern number can written:

\be
C_2(\varepsilon_F) = \frac{1}{32\pi^2}\int_{BZ} d\vect{k}\,\epsilon_{lmno}\textrm{Tr}[F_{lm}(\mathbf{k})F_{no}(\mathbf{k})],\label{eq:C2deg}
\ee

in terms of the non-Abelian Berry curvature $F^{\alpha\beta}_{lm}$ and where the trace is taken over the occupied bands. The above expression can be simplified by taking advantage of the symmetries of the Berry curvature, i.e. $F_{lm} = -F_{ml}$, to the form:
\be
\label{eq:chern2_3terms}
C_2(\varepsilon_F) = \frac{1}{4\pi^2}\int_{BZ} d\vect{k} \; \text{Tr}[F_{xy}F_{zw}+F_{wx}F_{zy}+F_{zx}F_{yw}].
\ee

%%%%%%%%%%%%%%%%%%%%%%%%%%%%%%%%%%%%
\section{Outline of the algorithm}
%%%%%%%%%%%%%%%%%%%%%%%%%%%%%%%%%%%%

In general, the second Chern number cannot be calculated analytically and one should therefore find an efficient algorithm to compute it numerically. A first attempt would be to define a lattice on the Brillouin zone and discretize derivatives of Eq.~\eqref{eq:current4D}. However, this strategy proves to be not efficient since this discretization is not gauge invariant and this method converges thus very slowly. We here propose the use of lattice gauge theory to circumvent this problem, as already emphasized for the computation of the first Chern number~\cite{Phillips_1985,Fukui_2005,resta2007theory}.   In this section, we briefly review lattice gauge theory and describe the different steps of the algorithm. To this end, we define the Berry curvature on a lattice, called field strength tensor, and discuss its convergence to the continuum. The latter allows one to compute the first Chern number in two dimensions and can be generalized for the computation of the second Chern number in four dimensions. Note that the non-Abelian lattice gauge theory approach can be also used for calculating other 
topological invariants, see for instance~\cite{Yu2011}.

\subsection{Review of the algorithm for the first Chern number}

Historically, the Chern number has been linked to the quantization of the transverse conductivity via the Kubo formula~\cite{Thouless_1982}. This formulation has also been generalized to systems with interactions or disorder by the introduction of twisted boundary conditions~\cite{Niu_1985}. This description is based on the evaluation of the velocity operator. The integral over the Brillouin zone can be written as a Riemann integral. However, the latter converges slowly~\cite{Dauphin_2015}. We here review an algorithm based on the discretized version of Wilson loops (a introductory review of the Wilson loop and its applications can be found in~\cite{Makeenko2010}). Let us consider the Brillouin zone and define a finite grid of $n_x\times n_y$ small plaquettes of size $\Delta\times \Delta$ at momentum $\mathbf{k}_l$, $\mathbf{k}_l$ being the lowest left corner of the plaquette. The total Chern number $C_1(\varepsilon_F)$ of Eq.~\eqref{eq:firstchernnum} can be written as the sum  over all the plaquettes of the grid:

\begin{equation}
\label{eq:totalchgrid}
C_1(\varepsilon_F)=\frac{1}{2\pi}\sum_{\{k_l\}} \text{Im}[\text{Tr}\left(P\,F_{xy,l}\,P\right)]\,\text{,}
\end{equation}

where $P=\sum_\text{occ. bands} \vert \psi^\alpha \ra \la \psi^\alpha \vert$ is the projector on the ground state of the occupied bands and 

\begin{equation}
F_{xy,l}=\int_\square F_{xy}(\mathbf{k}) d\mathbf{k}
\end{equation}

is the non-Abelian Berry curvature contribution on each plaquette. Let us discuss how to efficiently compute Eq.~\eqref{eq:totalchgrid}. To this end, we first define the link tensor:

\be
(U_\mu)^{\alpha\beta} (k)= \la \psi^\alpha(\vect{k})|\psi^\beta(\vect{k}+\Delta \,\mathbf{1}_\mu)\ra.
\ee

The latter describes the phase acquired during the parallel transport between two neighboring sites. This operator is unitary. Next, we {define the Wilson loop~\cite{kogut_1979,Asboth_2016,Alden_1992} around the plaquette by performing the product 
\be
U^P_{xy,l}=\Pi_\square U_\mu={U^P}_x(\vect{k}_l)\,U^P_y(\vect{k}_l+\hat{x})\,U^P_x(\vect{k}_l+\hat{y})^{-1}\,U^P_y(\vect{k}_l)^{-1}\text{,}
\ee

where 
\be
U^P_{\mu}(\vect{k})\equiv P\, U_\mu(\vect{k})\,P\,\text{.}
\ee

 If one performs Taylor expansion of the $U^P_{\mu}$ up to the second order in $\Delta$ and compute the product up to the second order in $\Delta$, one can show that $U^P_{xy,l}=1+F^P_{xy}(\mathbf{k}_l)\,\Delta^2+o(\Delta^2)$~\cite{kogut_1979,Alden_1992}, where $F^P_{xy}(\mathbf{k}_l)$ is the non-Abelian Berry connection over the occupied bands defined in the previous section. We therefore define the field-strength tensor as $\tilde{F}^P_{xy,l}=\ln  \, U^P_{xy,l} $. The lattice Chern number is then defined as
\be
\label{eq:algo:nonabch}
\tilde{C}(\varepsilon_F)=\frac{1}{2\pi}\sum_{\{\mathbf{k}_l\}}\text{Im}\left[\text{Tr} \, \tilde{F}^P_{xy,l}\right]\text{.}
\ee}

Finally, by using the identity $\text{Tr}[ \ln A]=\ln[\text{det}(A)]$ and one can then write Eq.~\eqref{eq:algo:nonabch} as

\begin{align}
\tilde{C}(\varepsilon_F)&=\frac{1}{2\pi}\sum_{\{\mathbf{k}_l\}}\text{Im}\left\{\ln \left[ \det \left(U^P_{x}(\vect{k}_l)\,
                                 U^P_{y}(\vect{k}_l+\hat{x})\,U^P_{x}(\vect{k}_l+\hat{y})^{-1}\,U^P_{y}(\vect{k}_l)^{-1} \right) \right]\right\}\cr
                        &=\frac{1}{2\pi}\sum_{\{\mathbf{k}_l\}}\text{Im}\left\{\ln \left[ \det \left(U^P_{x}(\vect{k}_l)\right)\,
                                \det \left( U^P_{y}(\vect{k}_l+\hat{x})\right)\,\det \left(U^P_{x}(\vect{k}_l+\hat{y})^{-1}\right)\,\det \left(U^P_{y}(\vect{k}_l)^{-1} \right) \right]\right\}         
\text{.}
\end{align}

The quantity above is manifestly gauge invariant, where we admit as gauge transformations  any {unitary rotation of the basis of the occupied states}. However, it is not generically an integer number. In principle, for a generic choice of the lattice in momentum space, $\tilde C$  converges {\it at worst} as $1/\Delta$ where $\Delta$ is the lattice spacing in momentum space, but it will in general converge faster, see our numerical results of Sec.~\ref{sec:benchmark}. In fact, we notice that if we take care of not including in our lattice the momenta where the bands become degenerate (see also Ref.~\cite{Hatsugai_2006}), we can treat the problem as Abelian and restrict the diagonal of the $U_p$'s. This essentially happens because $\text{Tr}(F) = \text{Tr}(dA)$, as the trace of a commutator vanishes and the first Chern number receives contribution only from the Abelian part. As the determinant of a diagonal matrix is the product of the diagonal elements, we can define the Abelian link as in Ref.~\cite{Fukui_2005} and the 1st Chern number can be taken as it would be the sum of 1st Chern numbers of the bands. In Ref.~\cite{Fukui_2005}, the authors defined the quantity to be summed as the normalized Abelian link (the Abelian link divided by its norm). By doing so, as shown in Ref.~\cite{Fukui_2005} this turns $\tilde C$ in a integer quantity, i.e. the number of cuts encountered in the logarithm that defines the link phase, which is fast converging with the lattice spacing $\Delta$. Again, the velocity of convergence depends on the choice of the discretization grid. In Ref.~\cite{Hatsugai_2006}, the authors generalized their algorithm to the case of degenerate bands.  They use the same construction as the one presented above. The only difference again is that they normalize the contribution from the non-Abelian link such that by construction the algorithm gives integer numbers. 

\subsection{Algorithm for the second Chern number}

We now propose an algorithm to compute the second Chern number $C_2$ of Eq.~\eqref{eq:chern2_3terms}. We consider the Brillouin zone and define a finite grid of $n_B\times n_B \times n_B \times n_B$ small hypercubes at momentum $\mathbf{k}_l$. The second Chern of the occupied bands number is then written as:

\be
C_2(\varepsilon_F)=\frac{1}{4\pi^2}\sum_{\{\mathbf{k}_l\}}\text{Tr}[P\, F_l \, P]\text{,} \label{eq:2ndplaquette}
\ee
where 
\be
\label{eq:deffl}
F_l=\int_\square \left[F_{xy}(\mathbf{k}_l)PF_{zw}(\mathbf{k}_l)+F_{wx}(\mathbf{k}_l)PF_{zy}(\mathbf{k}_l)+F_{zx}(\mathbf{k}_l)PF_{yw}(\mathbf{k}_l)\right] \, d^4k
\ee
is integrated over the hypercube. {As discussed in the previous section,  one can approximate the Berry curvature by using the field strength tensor obtained from a plaquette of side $\Delta$ with an error of $o(\Delta^2)$. Thus, we also propose here to compute the Berry curvature with the help of the Wilson loop

\be
U^P_{\mu\nu,l}\equiv U^P_\mu(\mathbf{k}_l)\,U^P_{\nu}(\vect{k}_l+\hat{\mu})\,U^P_{\mu}(\vect{k}_l+\hat{\nu})^{-1}\,U^P_{\nu}(\vect{k}_l)^{-1} \label{eq:approx2}
\ee

and use it to calculate the contribution to the second Chern number of each hypercube cell of the hyperlattice in momentum space. Therefore, in practice, we compute the approximant of the second Chern $\tilde{C}_2$ number using  the following equation 

\be
\tilde{C}_2(\varepsilon_F)=\frac{1}{4\pi^2}\sum_{\{\mathbf{k}_l\}}\text{Tr}\, \tilde{F}^P_l\text{,} 
\ee

where

\be
\label{eq:deffltilde}
\tilde{F}^P_l= \tilde{F}^P_{xy}(\mathbf{k}_l)\tilde{F}^P_{zw}(\mathbf{k}_l)+\tilde{F}^P_{wx}(\mathbf{k}_l)\tilde{F}^P_{zy}(\mathbf{k}_l)+\tilde{F}^P_{zx}(\mathbf{k}_l)\tilde{F}^P_{yw}(\mathbf{k}_l)
\ee

is defined in terms of

\be
\tilde{F}^P_{\mu\nu}(\mathbf{k}_l)=\ln( U^P_{\mu\nu,l}).
\ee}

 Let us make some comments
\begin{itemize}
\item As already discussed in the previous section, the algorithm converges for a generic model and a generic lattice in momentum space as $\tilde C_2(\epsilon_F)-C_2(\epsilon_F)\sim 1/\Delta$,
where $\Delta=2\pi/N$ is the lattice spacing in momentum space and with $N$ the number of points on the side of our hypercubic. In fact, we typically observe much better 
behavior. In order to estimate the ``typical'' convergence one should evaluate which is the error computed on average for different position of the lattice grid. Such calculation is 
beyond the scope of the present work;\\
\item As noticed above, it is in general not possible, to the best of our knowledge, to ``Abelianize'' the calculation of the second Chern number {in the case of intersecting bands}, 
as the commutators of the Berry connection are not cancelling out of the expression of Eq.~\eqref{eq:C2deg}. In addition, one cannot ``exchange'' the trace with the determinant  and thus
get an approximant of the second Chern $\tilde C_2$ that has integer values;\\

\end{itemize}

We finally conclude this section by comparing this method to the one introduced in Ref.~\cite{Price_2015}. In this work, the authors introduced an algorithm, a generalization of the algorithm of Ref.~\cite{Fukui_2005}, to compute the second Chern number of non-intersecting bands and relies on the Abelian version of the Berry curvature. As discussed previously, in the presence of intersecting bands, one cannot avoid the use of a non-Abelian links for calculating the second Chern number.

%%%%%%%%%%%%%%%%%%
\section{Benchmark, peformance and application of the algorithm} \label{sec:benchmark} 
%%%%%%%%%%%%%%%%%%

We now benchmark the algorithm on several examples, characterize the convergence of the algorithm in these particular cases. The first model is the lattice Dirac model, a lattice version of the continuous 4D Dirac model. This model is exactly solvable and the second Chern number can be written as an analytical integral. The second model is the 4D quantum Hall effect with two perpendicular magnetic fluxes, for which quantum simulations with cold atoms and photons have been proposed recently. In this case, the system should be treated numerically. Here, an identity allows us to write the second Chern number as a sum of product of first Chern numbers. The last model is the 4D quantum Hall effect with coupled fluxes. In this case, it is not possible to decouple the model into two 2D models.

%-------------------------------------------------------------------------------------------------------------------------------
\subsection{Lattice Dirac model}
%-------------------------------------------------------------------------------------------------------------------------------
We first benchmark the algorithm on the lattice Dirac model in $(4+1)$  dimensions~\cite{Qi_2008}. In this model, the second Chern number can be written in terms of the Poyntryagin index and computed analytically. This is thus a good candidate to test the algorithm and compare the results with exact values of the second Chern number.

\begin{figure}[h]
\includegraphics[scale=0.45]{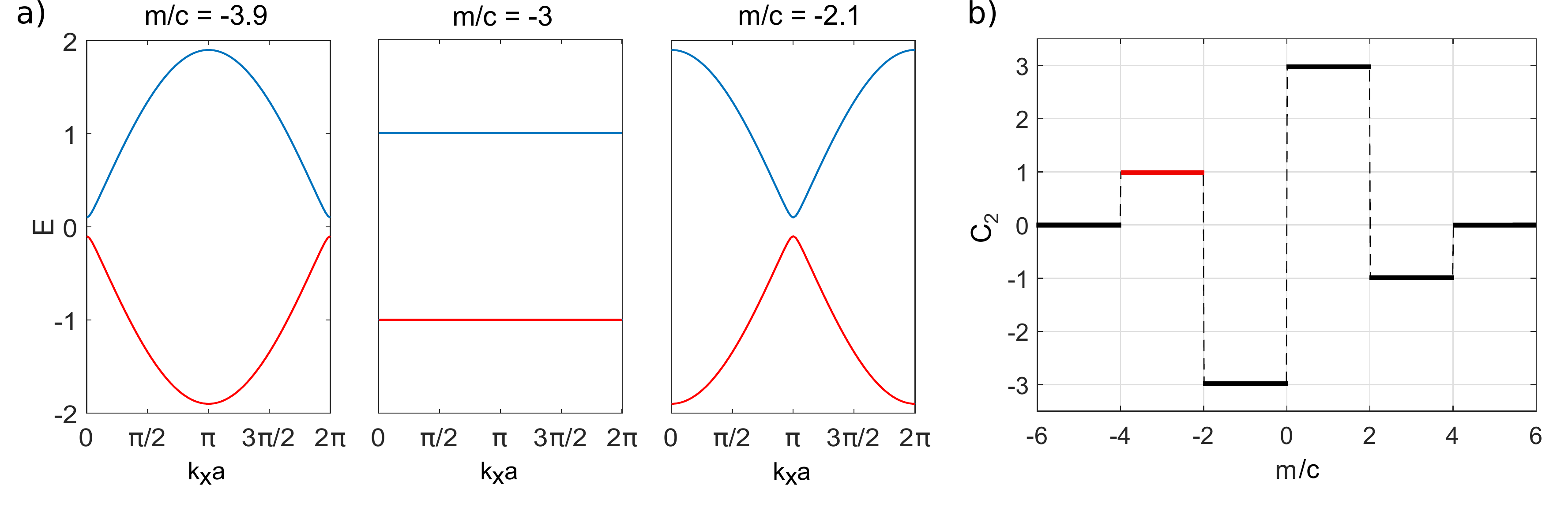}
\caption{\textbf{a.} Energy spectrum of the 4D lattice Dirac model for three values of the parameter $m$. The nearest critical points are for $m = -4$ and $m = -2$ and one can observe reduction of the band gap in their the vicinity (figures for $m = 3.9$ and $m = 2.1$). \textbf{b.} Phase diagram for the $(4+1)$-dimensional Dirac model for the parameter $c=1$. The interval of the $m$ corresponding to the panel a) is marked on the phase diagram by the red colour.}
\label{Dirac_phase}
\end{figure}

In the continuum limit, the Dirac Hamiltonian in $(4+1)$ dimensions takes the form
\be
H_{Dirac} = \int{d^4 x [\psi^\dagger(x)\Gamma^j(-i\partial_j)\psi(x) + m\psi^\dagger\Gamma^0\psi]},
\ee
where $j = 1,2,3,4$ are the spatial dimensions and $\Gamma^\mu$, $\mu = 0,1,2,3,4$, are Dirac matrices satisfying the anticommutation relations $\{\Gamma^\mu,\Gamma^\nu\} = 2\delta_{\mu\nu}\mathbb{I}$. The lattice version of the Dirac model in tight-binding limit~\cite{Qi_2008} is given by 
\be
H_{latt} = \sum_{n,j}[\psi^\dagger_n\left(\frac{c\Gamma^0-i\Gamma^j}{2}\right)\psi_{n+\hat{j}} + H.c.]+m\sum_n\psi^\dagger_n\Gamma^0\psi_n,
\ee 
where  the Dirac matrices $\Gamma = (\sigma_x\otimes\mathbb{I},\sigma_y\otimes\mathbb{I},\sigma_y\otimes\sigma_x,\sigma_y\otimes\sigma_y,\sigma_z\otimes\sigma_z)$ are written as tensor products of Pauli matrices $\sigma_\mu$. In the momentum space, the lattice Dirac Hamiltonian takes the form
\be
H_{latt} = \sum_\vect{k}\psi^\dagger_\vect{k}d_a(\vect{k})\Gamma^a\psi_\vect{k},
\ee
where $\mathbf{d}(\vect{k})=\left(\left(m+c\sum_j\textrm{cos}k_j\right),\textrm{sin}k_x,\textrm{sin}k_y,\textrm{sin}k_z,\textrm{sin}k_w\right)$. The energy spectrum can be found analytically: the system has two degenerate energy bands $E_+(k)$ and $E_-(k)$ and is shown for different values of $m$ in Fig.~\ref{Dirac_phase}. For this particular model, the second Chern number can be mapped to the Poyntargyn number~\cite{Qi_2008}
\be
C_2 = \frac{3}{8\pi^2}\int d^4k\epsilon^{abcde}\hat{d}_a\;\partial_x\hat{d}_b\;\partial_y\hat{d}_c\;\partial_z\hat{d}_d\; \partial_w\hat{d}_e,
\label{2cn_integral}
\ee
where $\hat{d}_a(\vect{k})\equiv d_a(\vect{k})/|d(\vect{k})|$ and $a,b,c,d,e = 0,1,2,3,4$. It turns out that the integral of Eq.~\eqref{2cn_integral} has analytical solution that depends on the parameters $m$ and $c$, as shown in Fig. \ref{Dirac_phase}. 

%%%%%%%%%%%%%%%%%%%%%%%%%%%%%%%%%%%%%%%%%%%%%%%%%%%%%%%%%%%%%%%%%%%%%%%%%%%
%%%%%%%%%%%%%%%%%%%%%%%%%%%%%%%%%%%%%%%%%%%%%%%%%%%%%%%%%%%%%%%%%%%%%%%%%%%

\begin{figure}[h]
%\hspace{-1.48cm}
\includegraphics[scale=0.5]{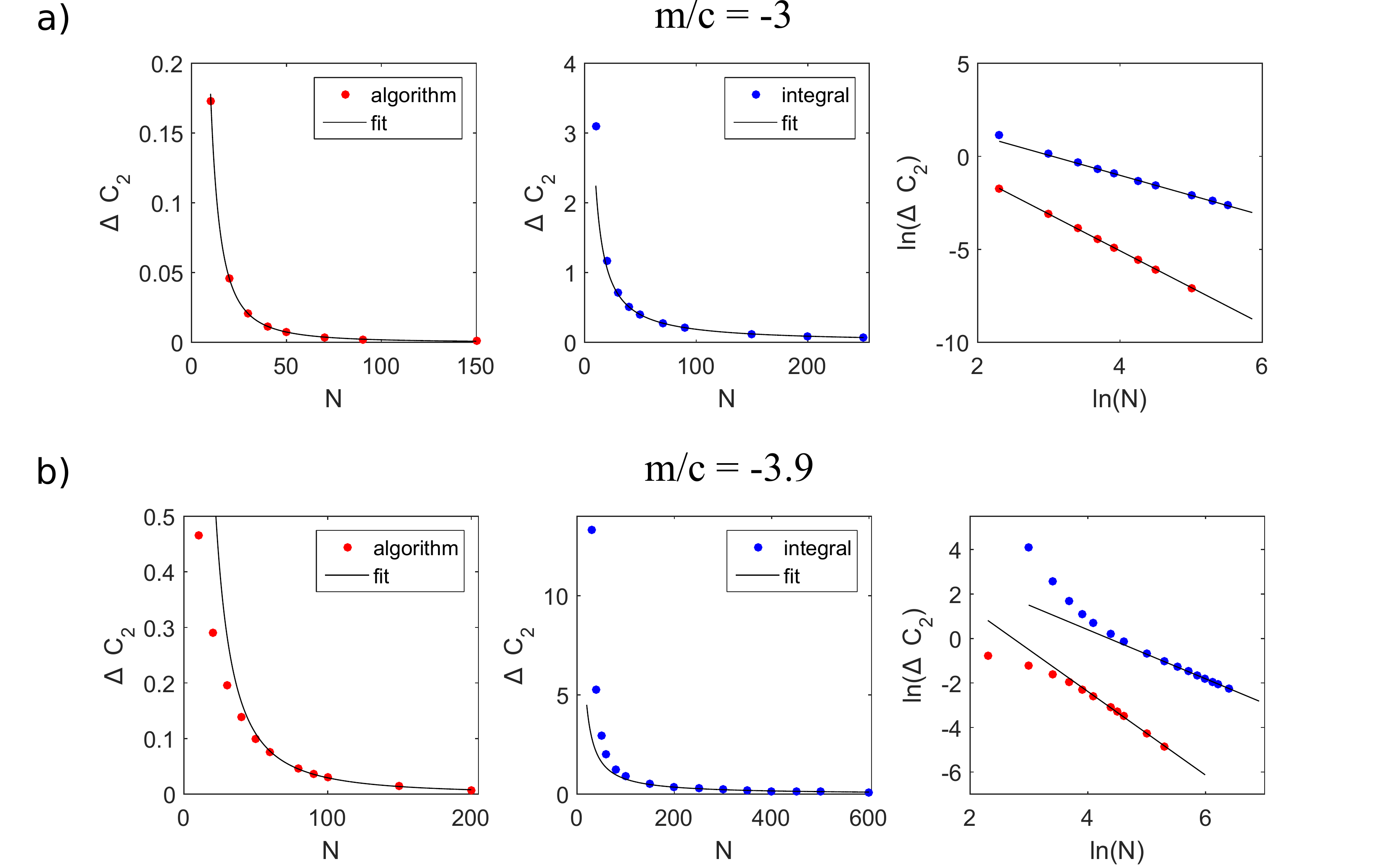}
\caption{Plots show $\Delta C_2$, i.e. the difference between the exact value of the second Chern number and the numerical result, as the function of the number of points  $N$ in the Brillouin zone. Upper panel corresponds to $m = -3$, i.e. far from the transition, so that the convergence regime is obtained even for coarse grid of the Brillouin zone. The situation is different for $m=-3.9$ (bottom panel) near the transition. For our method the convergence regime is obtaind for approximately $N = 60$, while the integral formulation needs the number of points around $N = 250$.}
\label{Dirac_conv}
\end{figure}

We benchmark the algorithm for several values of $m$ and compare the difference to the exact value of the second Chern number $\Delta C_2$. Results are shown in Fig.~ \ref{Dirac_conv}a for $m=-3$ in terms of the number of grid points in each directions $N$. The results of the algorithm (red points) are compared to the Riemann sum of the integral of Eq.~\eqref{2cn_integral}. The convergence is found by taking the log-log plot of the graphs. One finds with the help of linear regression that the algorithm converges in $N^{-2}$ and the Riemann sum in $N^{-1}$, showing the fast convergence of the algorithm. Figure~\ref{Dirac_conv}b shows the convergence for $m=-3.9$, close to the gap closing. In this case, one can distinguish two regimes: before the convergence and the convergence. As expected, the algorithm quickly reaches convergence around $N=60$, much before the discretization of the integral which appears around $N=250$.

%%%%%%%%%%%%%%%%%%%%%%%%%%%%%%%%%%%%%%%%%%%%%%%%%%%%%%%%%%%%%%%%%%%%%%%%%%%
%%%%%%%%%%%%%%%%%%%%%%%%%%%%%%%%%%%%%%%%%%%%%%%%%%%%%%%%%%%%%%%%%%%%%%%%%%%

%-------------------------------------------------------------------------------------------------------------------------------
\subsection{4D quantum Hall effect}
%-------------------------------------------------------------------------------------------------------------------------------

%%%%%%%%%%%%%%%%%%%%%%%%%%%%%%%%%%%%%%%%%%%%%%%%%%%%%%%%%%%%%%%%%%%%%%%%%%%
\begin{figure}[h]
%\hspace{-1.48cm}
\includegraphics[scale=0.4]{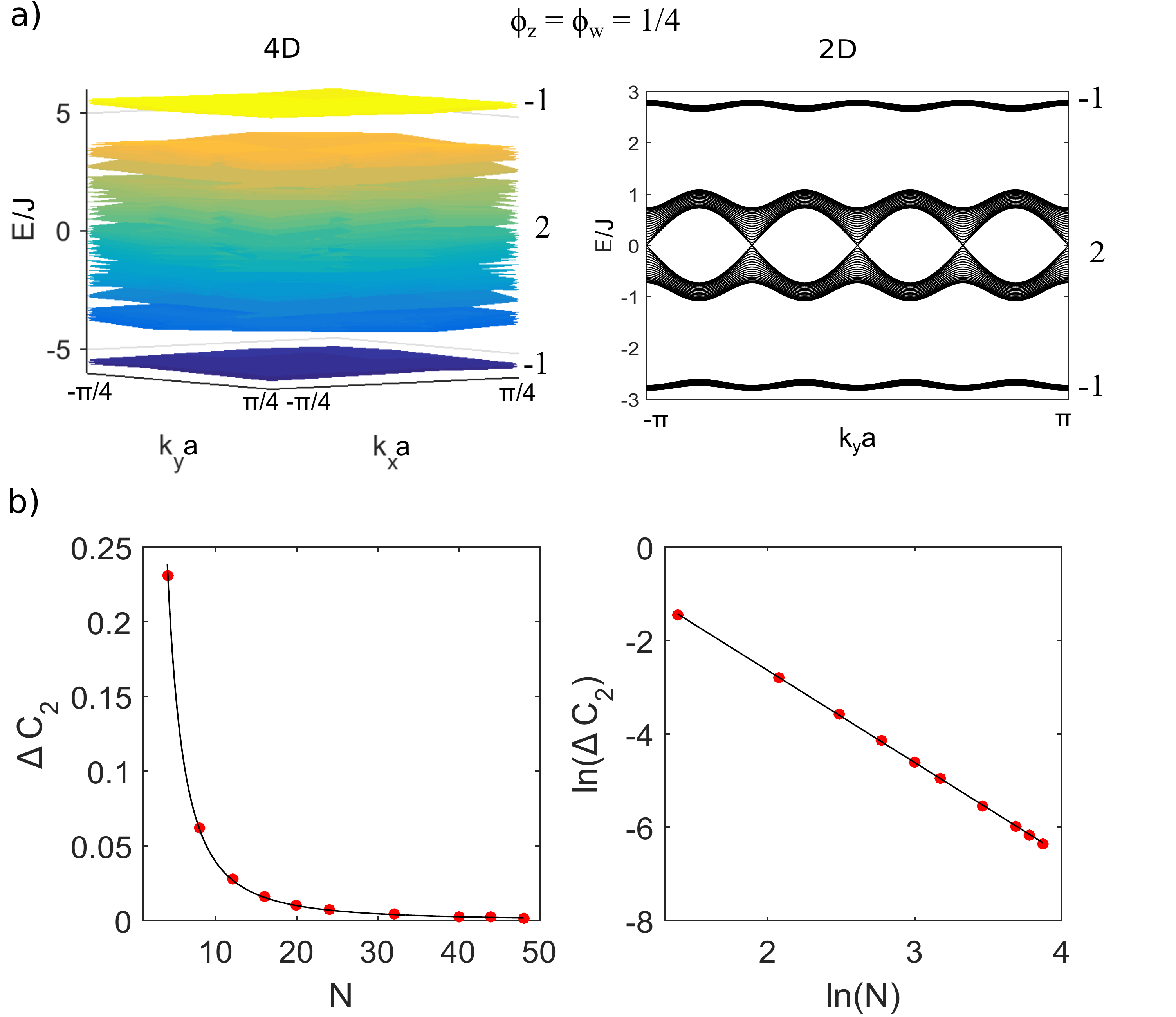}
\caption{a) Energy spectrum of the 4D QHE, see Eq. (\ref{Harper_4D}), that occures as the Minkowski sum of the two uncoupled 2D models (right panel) for the flux $\phi_z = \phi_w = 1/4$ with the corresponding values of the 2CN on the right hand side. b) Plots show $\Delta c_2$, i.e. the difference between the exact value of the 2CN and the numerical result, as the function of the number of points in the Brillouin zone $N$. Red dots correspond to the numerical values calculated by our algorithm. The parameters of the linear fit $\textrm{ln}(\Delta c_2) = a\cdot\textrm{ln}(N)+b$ are $a = -1.973\pm0.018$, $b = 1.303\pm0.056$ with the accuracy $R^2 = 0.9999$.}
\label{qhe_14}
\end{figure} 
%%%%%%%%%%%%%%%%%%%%%%%%%%%%%%%%%%%%%%%%%%%%%%%%%%%%%%%%%%%%%%%%%%%%%%%%%%%

We now benchmark the algorithm on the 4D quantum Hall effect~\cite{Zhang_2001}. This model is experimentally relevant since the recent implementation schemes have been proposed  for optical lattices with synthetic dimensions~\cite{Price_2015} and for quasi-crystals~\cite{Kraus_2013}. The 4D quantum Hall effect has been recently experimentally observed in a 2D Thouless pump experiment~\cite{Lohse2018,Zilberberg2018}. The model is described by a tight-binding Hamiltonian of spinless fermions:

\be
H = - \sum_{{\bf r},\hat{\mu}}J  c_{\bf r}^\dagger e^{i A_{\hat{\mu}}({\bf r}) } c_{{\bf r}+\hat{\mu}} + H.c.,
\label{hamiltonian}
\ee
where ${\bf r} = (x,y,z,w), \hat{\mu} $ is the nearest neighbor link, the operator ${\bf c}_{\bf r}^{\dagger}$ creates a particle at site ${\bf r}$. The gauge is chosen such that $A_x({\bf r}) = A_y({\bf r}) = 0$, $A_z({\bf r}) = 2 \pi \phi_z x $ and $A_w({\bf r}) = 2 \pi \phi_w y$. For $\phi_z=p_z/q_z$ and $\phi_w=p_w/q_w$, the system can be written in the momentum space by considering a magnetic unit cell of $q_z$ sites in the x-direction and $q_w$ sites in the y-direction.The generalized Harper equation reads then:
\bea
H (\mathbf{k}) =&-J\left(e^{ik_x}u_{m+1,n}(\vect{k}) + e^{-ik_x}u_{m-1,n}(\vect{k})\right) - J\left(e^{ik_y}u_{m,n+1}(\mathbf{k})+e^{-ik_y}u_{m,n-1}(\mathbf{k})\right)  \nonumber \\ 
&-2Ju_{m,n}(\vect{k})(\textrm{cos}(2\pi\phi_z m + k_z) + \textrm{cos}(2\pi\phi_w n + k_w)) = \varepsilon(\vect{k})u_{m,n}(\vect{k}),
\label{Harper_4D}
\eea
where $x = ma$ and $y = na$. While the system is genuine 4D as the particle can hop in all 4D dimension, the Harper equation can be rewritten in terms of two uncoupled Harper equations corresponding to two 2D Hofstadter models:

\be
\begin{split}
H (\mathbf{k}) =&-J[e^{ik_x}u_{m+1,n}(\vect{k}) + e^{-ik_x}u_{m-1,n}(\vect{k})]-2Jv_{m,n}(\vect{k})\textrm{cos}(2\pi\phi_z m + k_z)=\varepsilon_{xz}(\vect{k})u_{m,n}(\vect{k})\\ 
&- J[e^{ik_y}v_{m,n+1}(\mathbf{k})+e^{-ik_y}v_{m,n-1}(\mathbf{k})]-2Ju_{m,n}(\vect{k}) \textrm{cos}(2\pi\phi_w n + k_w) = \varepsilon_{yw}(\vect{k})v_{m,n}(\vect{k})
\end{split}
\ee

 and the energy spectrum can be written as a Minkowski sum of its two 2D Hofstadter models $\varepsilon(\vect{k}) = \{\varepsilon_{xz}(\mathbf{k}) + \varepsilon_{yw}(\mathbf{k})\}$~\cite{Kraus_2013,Price_2015}. Figure \ref{qhe_14}a shows an example the 4D energy spectrum $\phi_z = \phi_w = 1/4$. The system has two non degenerate energy bands and one degenerate energy band (composed of 14 energy bands). The second Chern numbers can be computed from the first Chern number of the two decoupled models with the help of the identity:
 
 \be
 C_2(\varepsilon_F)=\sum_{\{\alpha,\beta \vert \varepsilon^{\alpha}_{xz}+\varepsilon^\beta_{yw}<\varepsilon_F\}} C^\alpha_{1,zx}C^\beta_{1,yw}.
 \ee
 
The lowest energy is non-degenerate. The algorithm is converging rapidly to the integer value even for the coarse grid of $3 \times 3 \times 12 \times 12$ sites and is thus converging as fast as the algorithm introduced in Ref.~\cite{Price_2015} to compute the second Chern number for non degenerate bands. The second energy band is highly degenerated and is composed of 14 energy bands. Results are presented in Fig.~ \ref{qhe_14}b. Left subfigure shows the relative error to the exact value in terms of the number of points in the Brillouin zone $N$. Right Subfigure shows the log log plot of the relative error. One can see that even for a coarse grid, the algorithm is in the convergence regime. The convergence in terms of N is found by taking the linear regression of the log log plot and one finds a convergence in $1/N^2$. Finally, we also consider the magnetic fluxes $\phi_w=\phi_z=3/5$. In this case, the first energy band is degenerate and is composed of 4 energy bands. Applying the algorithm, we find the convergence even for the coarse grid of $3\times 3 \times 15 \times 15$ points.

%-------------------------------------------------------------------------------------------------------------------------------
\subsection{Analysis of the system with coupled fluxes}
%-------------------------------------------------------------------------------------------------------------------------------

%%%%%%%%%%%%%%%%%%%%%%%%%%%%%%%%%%%%%%%%%%%%%%%%%%%%%%%%%%%%%%%%%%%%%%%%%%%
\begin{figure}[h]
%\hspace{-1.48cm}
\includegraphics[scale=0.45]{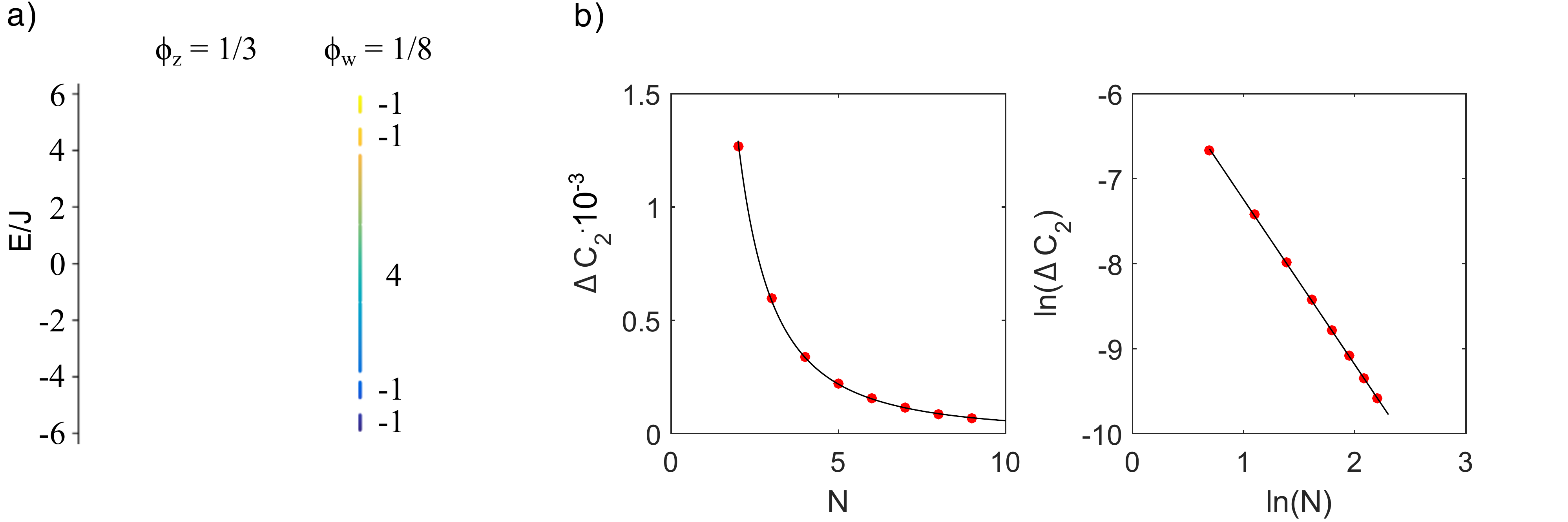}
\caption{a) Spectrum of the modified 4D model, see Eq. (\ref{Harper_4D_mix}), in which all coordinates are coupled by the magnetic fluxes $\phi_z = 1/3$ and $\phi_w = 1/8$. There are four gaps that are large enough to reveal Hall physics. On the right hand side of the figure one can find the values of 2CN for each band. b) Plots show $\Delta c_2$, i.e. the difference between the exact value of the 2CN and the numerical result, as the function of the number of points in the Brillouin zone $N$. Red dots correspond to the numerical values calculated by our algorithm. The parameters of the linear fit $\textrm{ln}(\Delta c_2) = a\cdot\textrm{ln}(N)+b$ are: $a = -1.942\pm 0.023$, $b = -5.305\pm0.037$ and $R^2 = 0.9999$.}
\label{mixed}
\end{figure} 
%%%%%%%%%%%%%%%%%%%%%%%%%%%%%%%%%%%%%%%%%%%%%%%%%%%%%%%%%%%%%%%%%%%%%%%%%%%

We finally consider a more complex vector potential for Eq.~\eqref{hamiltonian}: $A_x({\bf r}) = A_y({\bf r}) = 0$, $A_z(\vect{r}) = 2\pi\phi_z x$ and and $A_w(\vect{r}) = 2\pi\phi_w(y + x)$, with $\phi_z = p_1/q_1$ and $\phi_w = p_2/q_2$. In this case, the system cannot be anymore separated in two independent Harper equations and one thus need an eficient algorithm to compute numerically the second Chern number. Here, the generalized Harper equation is given by:
\bea
H (\mathbf{k}) =&-J\left(e^{ik_x}u_{m+1,n}(\vect{k}) + e^{-ik_x}u_{m-1,n}(\vect{k})\right) - J\left(e^{ik_y}u_{m,n+1}+e^{-ik_y}u_{m,n-1}\right)  \nonumber \\ 
&-2Ju_{m,n}(\vect{k})(\textrm{cos}(2\pi\phi_z m + k_z) + \textrm{cos}(2\pi\phi_w (n+m) + k_w)) = E(\vect{k})u_{m,n}(\vect{k}),
\label{Harper_4D_mix}
\eea
where $m =x/a$ and $n= y/a$. Due to the existence of the additional flux depending on $x$, the magnetic cell in the $x$ direction has to be enlarged with respect to the flux $\phi_w$, i.e. the periodicity in the $x$ direction equals to the least common multiple (LCM) of $q_1$ and $q_2$ $\text{LCM}(q_1,q_2)$. As the consequence the Brillouin zone in $k_x$ should be reduced to $0\le k_x\le 2\pi/\text{LCM}(q_1,q_2)$. For the numerical simulations, we work with $\phi_z = 1/3$ and $\phi_w = 1/8$ such that the first energy band is degenerate, as shown in Fig.~\ref{mixed}a. Results of the algorithm are presented in Fig.~\ref{mixed}b. Even for a coarse grid, the algortihm already converges with an error of the order of $10^{-3}$. The linear fit on the Log log plot gives a convergence rate close to $1/N^2$.

%%%%%%%%%%%%%%%%%%%%%%%%%%%%%%%%%%%%%%%
\section{Conclusions and outlook}
%%%%%%%%%%%%%%%%%%%%%%%%%%%%%%%%%%%%%%%
In this work, we constructed an efficient algorithm to compute the second Chern number. The algorithm is based on the lattice gauge theory and converges quickly. We discussed the different steps of the algorithm and characterized its convergence to the continuous version. We have also applied this method to several paradigmatic 4D topological insulators including the lattice Dirac model, the 4D quantum Hall effect and the generalized 4D quantum Hall effect. 

The development of observables and tools to characterize topological models in dimensions higher than three is timely and of growing importance, in view of the recent progress in quantum simulation. 
Indeed, quantum simulators of 4D quantum Hall effect have been proposed using synthetic dimensions or quasi-crystals, and other intriguing phenomena like the 5D generalization of Weyl semimetals~\cite{Lian2016} have been recently predicted. The first observations of the quadratic responses experimentally achieved through Thouless pump processes in 2D quantum systems. An efficient algorithm to compute the second Chern number is thus particularly important to experimentally identify the signature of the 4D quantum Hall effect. Furthermore, in order to simulate the generalized 4D quantum Hall effect one needs to determine the second Chern number with the help of this algorithm since in this case it can neither be computed analytically nor related to the first Chern numbers.

A direct generalization of the algorithm would be the study of the transverse conductivity with partially filled bands~\cite{Dauphin_2014}.  When the Fermi energy lies in an energy band, it is still possible to distinguish the contribution from the geometrical phase to the conductivity, often called Berry conductivity. The latter can be dominant in two dimensions in the so called anomalous Hall effect~\cite{Nagaosa_2010}. It would be interesting to generalize this notion to four dimensions and to propose an algorithm to compute the contribution of the second Chern number to the Berry conductivity.  Another outlook would be to generalize this algorithm to higher dimensional space such as the computation of the third Chern number in the 6D quantum Hall effect~\cite{Petrides_2018}. 

The exploration of topological and many-body quantum systems has just begun. Most of the studies as we do in this work, have focused up to now on just  one class of the periodic table in four dimensions. We expect that algorithms for computing the topological invariants and for characterizing topological models in other classes and in four and higher dimensions will be needed soon.

%%%%%%%%%%%%%%%%%%%%%%%%%%%%%%%%%%%%%%%
\section*{Acknowledgments}
%%%%%%%%%%%%%%%%%%%%%%%%%%%%%%%%%%%%%%%
We thank H. Price and P. Massignan for a careful reading of the manuscript and insightful comments. 
MMG is supported by National Science Centre via project numbers DEC-2012/05/N/ST2/02745 and DEC-2014/12/T/ST2/00325. 
AD, ML acknowledge financial support from Spanish MINECO (Severo Ochoa SEV-2015-0522 and FisicaTeAMO FIS2016- 79508-P), the Generalitat de Catalunya (SGR 874 and CERCA), Fundacio Privada Cellex, and EU grants OSYRIS (ERC-2013-AdG Grant 339106), QUIC (H2020- FETProAct-2014 641122), and Fundaci\'o Cellex. AD is financed by a Cellex-ICFO-MPQ fellowship.
AC acknowledges financial support from the ERC Synergy Grant UQUAM and the SFB FoQuS (FWF Project No. F4016-N23).

\bibliography{bibliography}

\end{document}